\documentclass[a4paper, amsfonts, amssymb, amsmath, reprint, showkeys, nofootinbib, twoside]{revtex4-1}
\usepackage[english]{babel}
\usepackage[utf8]{inputenc}
\usepackage[colorinlistoftodos, color=green!40, prependcaption]{todonotes}
\usepackage{amsthm}
\usepackage{mathtools}
\usepackage{physics}
\usepackage{xcolor}
\usepackage{graphicx}
\usepackage[left=23mm,right=13mm,top=35mm,columnsep=15pt]{geometry} 
\usepackage{adjustbox}
\usepackage{placeins}
\usepackage[T1]{fontenc}
\usepackage{lipsum}
\usepackage{csquotes}
\usepackage[pdftex, pdftitle={Article}, pdfauthor={Author}]{hyperref} 
\begin{document}

\title{Tunable Wave Propagation Bandgap Via Stretching kirigami Sheets}

\author{Hesameddin Khosravi} 
\email[Correspondence email address: ]{hkhosra@clemson.edu}

\author{Suyi Li}
\affiliation{Department of Mechanical Engineering, Clemson University, Clemson, SC, USA}

\begin{abstract}
This study examines the Braggs bandgap and its mechanical tuning in a stretch-buckled kirigami sheet with ``zig-zag'' distributed parallel cuts.  When stretched beyond a critical threshold, the kirigami buckles out-of-plane and generates a 3D periodic architecture.  Our theoretical calculation, numerical simulation, and experiments confirm the transverse elastic wave propagation bandgaps and their correlation to stretching.  This result opens an avenue of using kirigami as a simple and effective approach for creating and adapting periodicity for wave propagation control. 
\end{abstract}

\maketitle

The emergence of mechanical metamaterials --- which derive their properties primarily from the underlying architecture rather than the constituent material --- have unleashed a new era of material design and functionalities.    Exploiting such architecture-property correlations has enabled many extreme and even ``unnatural'' responses \cite{Zadpoor2016, Bertoldi2017, Surjadi2019, Kadic2019}, and one of the most consequential examples is the Braggs wave propagation bandgap \cite{Li2004}.   The ability to manipulate propagating waves, in pre-determined frequency bands, via Braggs scattering in a periodic medium has enabled countless applications in noise mitigation \cite{Javid2016}, wave transmission control \cite{Lemoult2013}, acoustic cloaking \cite{Chen2007}, non-reciprocity \cite{Nassar2020}, and even wave-based mechanical computation \cite{Zangeneh2021}.   Nonetheless, the potentials of these applications all hinge upon our capability to fabricate and control periodic architecture in the metamaterial systems.  To this end, we have seen a variety of fabrication approaches such as 3D printing (e.g., Fused Deposition Modeling\cite{Kaur2017}, Direct Ink Writing \cite{Compton2014}, Electron Beam Melting \cite{Yang2015}), self-assembly \cite{Hayward2000}, and laser cutting \cite{Liu2018}.  However, many of them can be costly and time-consuming.  There is also a lack of simple and integrated approaches to physically change the underlying architecture on-demand, making it challenging to control the bandgap frequency in real-time. 

In this letter, we show that cutting and stretching elastic sheets using the kirigami principle is a simple, scalable, and effective approach for \emph{creating and controlling} periodic architectures, providing tunable wave propagation bandgaps.  kirigami (aka. ``cut-paper'') is an ancient art of transforming papers into sophisticated 3D shapes by cutting and then manipulation via folding and stretching.  It has recently transformed into a design and fabrication framework to engineer flexible electronics \cite{Song2015}, super-stretchable materials \cite{Chen2020}, light-weight structures \cite{Babaee2020}, and soft robots \cite{Khosravi2021, Cheng2020}.  Compared to other methods of constructing periodic 3D motifs, the kirigami-inspired approach has many unique advantages.  For example, the seemingly infinite possibilities in cutting pattern geometry offer significant freedoms in design.  And cutting as a fabrication method is easily scalable from nano- \cite{Chen2021}, micro- \cite{Chen2020b}, milli- \cite{Hwang2018}, to meter-scale \cite{Li2021}.  Nonetheless, the mechanical properties of kirigami-based structures and material systems remain an open research question.  While many static properties like auxetics \cite{Tang2017} and buckling \cite{Rafsanjani2017} have been examined, the kirigami structure's dynamic responses remain unexplored.

\begin{figure}[b]
    \centering
    \includegraphics[]{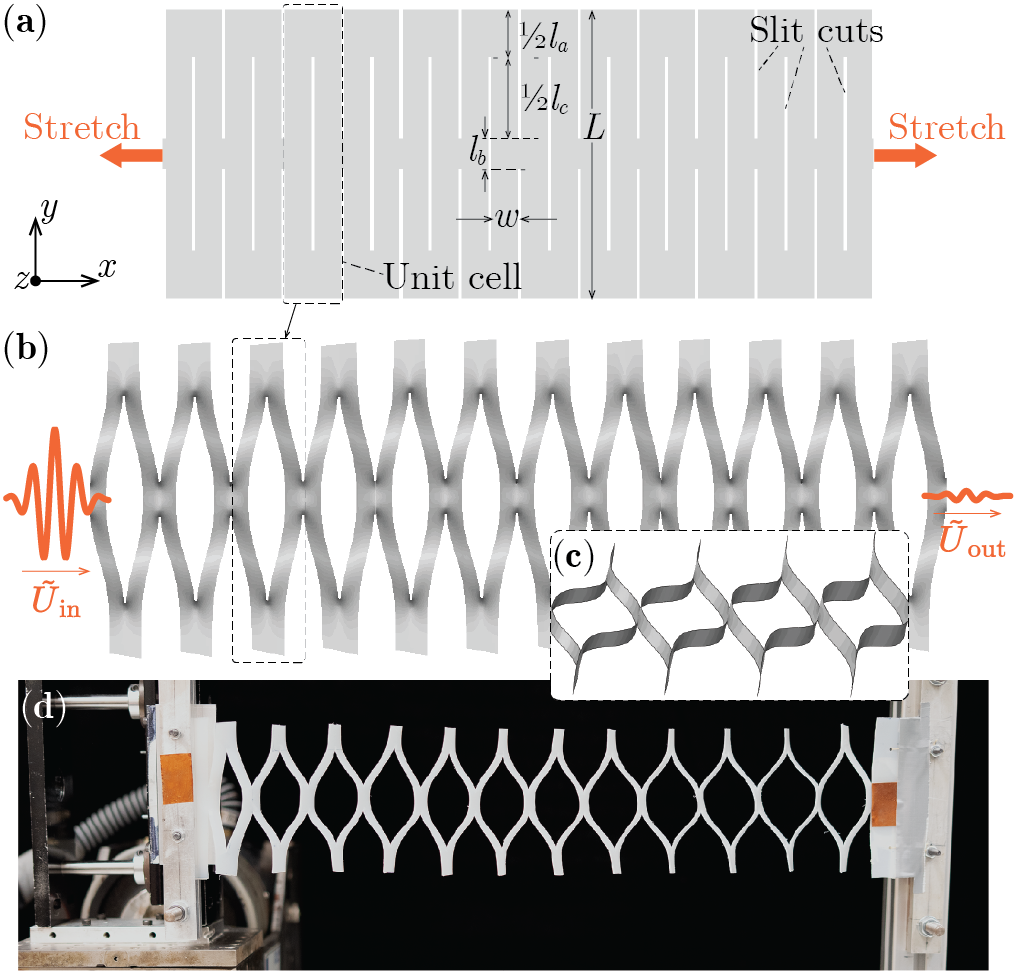}
    \vspace{-0.25in}
    \caption{The overall concept of this study. (a) A kirigami sheet with ``zig-zag'' patterned parallel cuts, whose geometry can be defined by just a few parameters.  Here, we neglect the slit cuts' width.  (b) The external shape of a stretched-buckled kirigami sheet, simulated in \texttt{ABAQUS} (3D Solid elements with Explicit Solver).  It exhibits a controllable, 3D periodic architecture, generating Braggs wave propagation bandgap. (c) A different view of the stretched kirigami sheet, showing the curved ligaments. (d) A kirigami prototype under testing.} 
    \label{fig:concept}
\end{figure}

\begin{figure*}[t]
   \centering
    \includegraphics[]{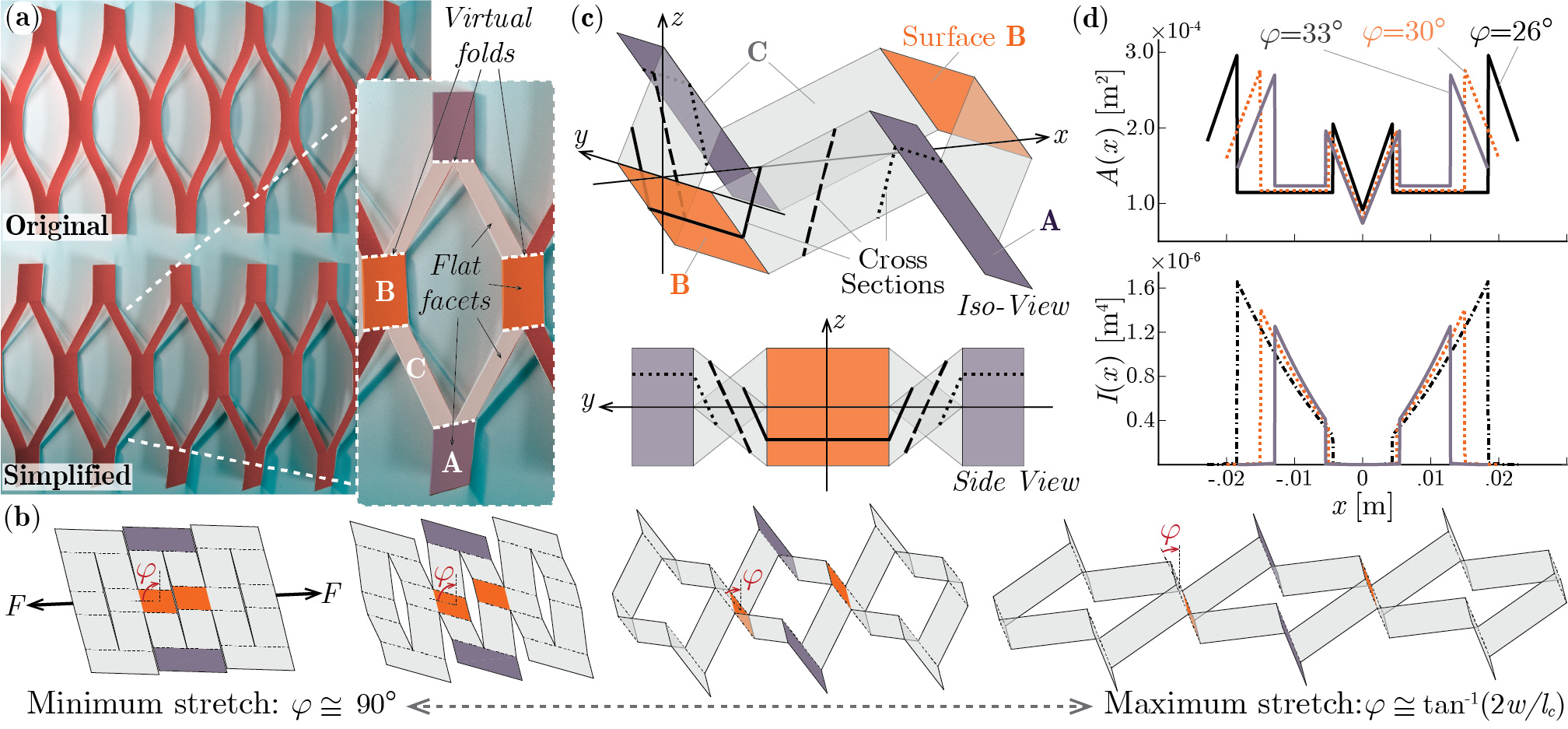}
    \caption{Simplification and modeling the geometry of stretch-buckled krigami sheet:  (a) Comparing the original and simplified geometry of kirigami sheets, the insert figure details the ``virtual folds'' and the flat facets (aka. Surface A, B, C) between them. (b) Stretching deformations of the simplified kirigami sheet, which shows rigid foldability, and its facets remain flat and undeformed during stretching. (c) Detailed geometry of a unit cell. The solid and dashed lines highlight the shape of its cross-sections.  (d) The $A(x)$ and $I(x)$ distribution in a unit cell with different stretches ($l_a=0.04$m, $l_b=0.02$m, $l_c=0.064$m, $w=0.01$m, $t=0.002$m). }
    \label{fig:geometry}
\end{figure*}

Here, we focus on the transverse elastic wave propagation in a stretch-buckled kirigami sheet with a ``zig-zag'' uniform distribution of parallel slit cuts shown in Figure \ref{fig:concept}.   A few design parameters --- including the cut size $l_a$, $l_b$, $l_c$, and the spacing between cuts $w$ --- can fully define the overall cutting pattern.  When sufficiently stretched in-plane along the periodicity direction (the $x-$direction), the kirigami sheet would buckle and develop out-of-plane deformations, creating a linear array of identical unit cells, each consisting of a few curved ligaments.  Moreover, controlling the in-plane stretching can directly change the 3D geometry of these unit cells, offering a simple approach for turning the dynamic responses.  The stretch-buckled kirigami sheet's overall height and width are significantly smaller than its length, and we assume its shearing deformation and rotational inertia are negligible.  So the Euler-Bernoulli beam condition applies, and the governing equation for the transverse wave propagating along the longitudinal $x-$direction is

\begin{equation} \label{eq:i1}
    \frac{\partial^2}{\partial x^2} \left[E I(x) \frac{\partial^2 {U(x,t)}}
    {\partial x^2}\right]+ \rho A(x)\frac{\partial^2 {U(x,t)}}{\partial t^2} = 0.
\end{equation}

\begin{figure*}[t]
   \centering
    \includegraphics[]{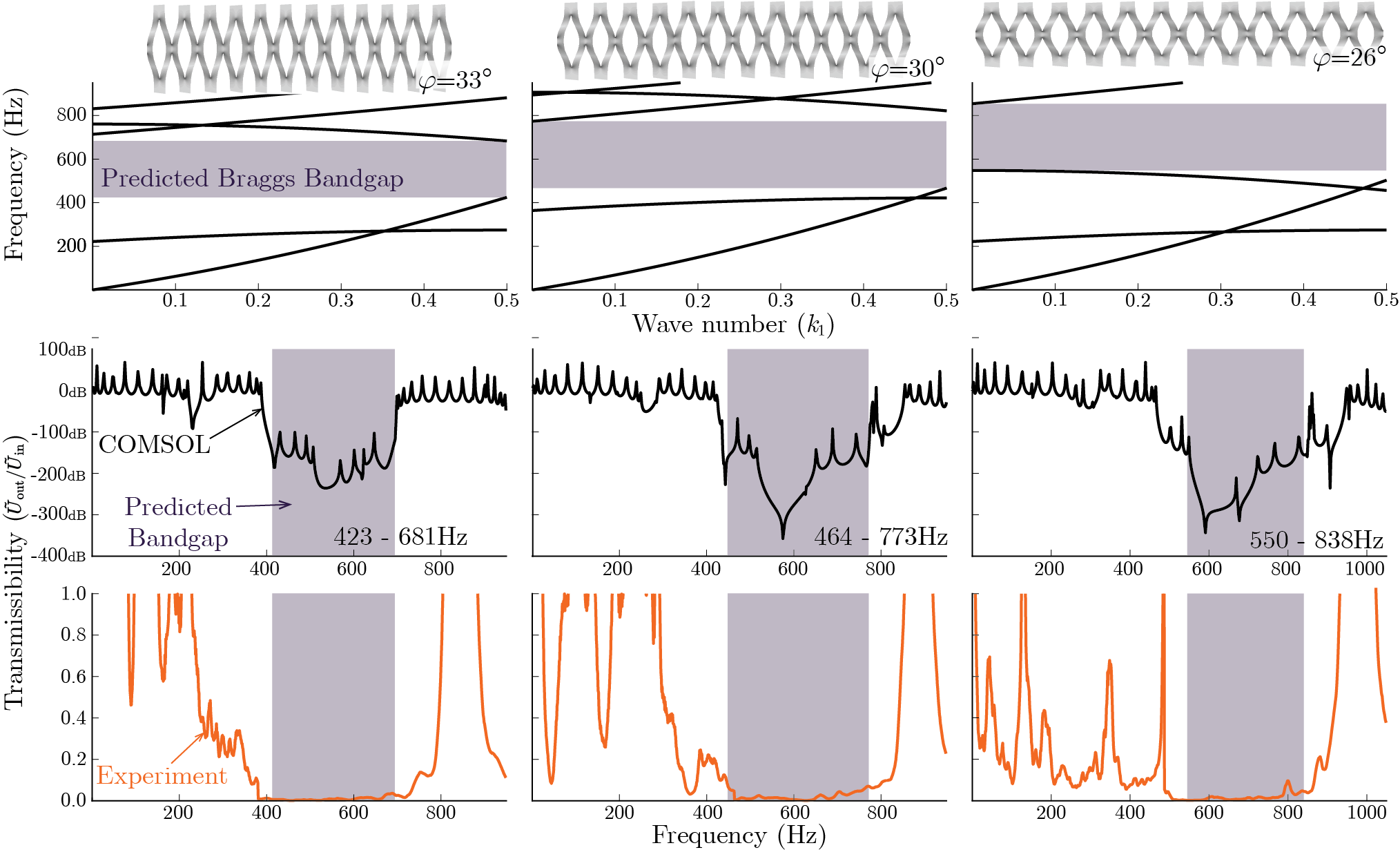}
    \vspace{-0.1in}
    \caption{Theoretically predicted dispersion relationships (first row), simulated transmissibility (second row), and experimentally measured transmissibility (third row) of a kirigami sample at three different levels of stretch. These results give consistent descriptions of Bragg's bandgap.}
    \label{fig:result}
\end{figure*}

Here, $\rho$ and $E$ are the mass density and Young's modulus of the constitutive sheet material, respectively. $U(x,t)$ is the out-of-plane transverse displacement in the $z-$direction.  $A(x)$ and $I(x)$ are the distribution of cross-section area and the second moment of inertia in the buckled kirigami sheet, respectively.  Calculating $A(x)$ and $I(x)$ is quite challenging because of the buckled kirigami sheet's complicated 3D geometry, especially the curved ligaments.  Therefore, we introduce ``virtual folds'' at critical locations on the ligaments that exhibit the most concentrated bending deformation and assume the facets in-between these folds are flat surfaces (Surfaces A, B, and C in Figure \ref{fig:geometry}).  Indeed, these folds can naturally occur when the stretching is strong enough to induce plastic deformation \cite{Rafsanjani2017}, so they could represent the mechanics of buckled kirigami sheets with reasonable accuracy.  Moreover, the simplified kirigami becomes rigid-foldable, making the kirigami deformation a one degree-of-freedom mechanism. 

Figure \ref{fig:geometry}(d) elucidates the complex $A(x)$ and $I(x)$ distribution in a unit cell subject to different amounts of in-plane stretch.  Here, we use the dihedral angle $\varphi$ between Surface B and $y-z$ reference plane to describe stretching, and the detailed calculations are available in Section 1 of the supplemental materials. 

Denote $\mathbf{R} = a_1 \mathbf{e}_1$ as a translational vector in the \emph{direct space}, representing the periodicity in the kirigami sheet. That is, we can construct the kirigami sheet by a set of infinite translational operations of the unit cell based on $\mathbf{R}$.   In this case, $a_1$ is an integer number, and the lattice vector $\mathbf{e}_1$ lies in the $x-$direction.  We then define the reciprocal lattice vector $\mathbf{G}$ in that $e^{-i\; \mathbf{G} \cdot \mathbf{R}}=1$.  Here, $m_1$ is another integer, $\mathbf{G}= m_1 \mathbf{b}_1$, and $\mathbf{b}_1$ is the lattice vector in the \emph{reciprocal space}.  Equivalent to the lattice vector $\mathbf{e}_1$ in the direct space, infinite translation operations of the reciprocal lattice vectors $\mathbf{b}_1$ defines the \emph{reciprocal space}, which has similar symmetry as the direct space.  Like in the direct lattice, an equivalent unit-cell can then be defined in the reciprocal space based on $\mathbf{b}_1$, giving us the First Brillouin Zone (FBZ).

To analyze the wave propagation behavior in the stretch-buckled kirigami sheet, we employ the plane wave expansion method (PWM).  First, a separation of variables gives $U(x,t)=\tilde{U}(x)e^{-i\omega t},$ where $\omega$ is the harmonic oscillation frequency.  Based on the Blotch theorem, we can formulate the spatial term $\tilde{U}(x)$ for the whole kirigami sheet into a product of plane-wave and periodic functions in the First Irreducible Brillouin Zone (FBZ) so that

\begin{equation} \label{eq:i3}
    \tilde{U}(x)= e^{i (k_1 b_1 x)} \sum_{n_1}{\hat{U}({n_1b_1})} e^{i(n_1 b_1 x)},
\end{equation}
where $b_1$ is the magnitude of reciprical lattice vector ($b_1=|\mathbf{b}_1|$), $n_1$ is an integer, and ${k}_1 $ is a wave number within the IBZ in that ${k}_1 \in [0,\frac{1}{2}]$. We further expand the distribution of cross-section area $A(x)$ and second moment of inertia $I(x)$ in Fourier series so that
\begin{align}
    I(x) & = \sum_{m_1}\hat{I}(m_1 b_1)e^{i(2\pi m_1 x)}\\
    A(x) & = \sum_{m_1}\hat{A}(m_1 b_1)e^{i(2\pi m_1 x)}
\end{align}

By substituting the formulations above into the governing equations of motion (\ref{eq:i1}), we obtain the eigenvalue problem for calculating the dispersion curves, and a few results are shown in the first row of Figure \ref{fig:result}.
\begin{multline} \label{eq:i5}
    \sum_{n_{1}} E(n_1+k_1)^2(k_1+m_1+n_1)^2\hat{I}(m_1 b_1) -\\ \omega^2\rho \hat{A}(m_1b_1)\;\hat{U} (n_1 b_1)= 0
\end{multline}

Our analytical theory's prediction, finite element simulations via COMSOL, and experimental tests confirm the occurrence of Bragg's bandgap in the stretch buckled kirigami sheet (Figure \ref{fig:result} and supplemental material details the simulation and experiment setup).   Moreover, the bandgap frequency can be effectively tuned by simply increasing or decreasing the in-plane stretch.  In these tests, we cut out the kirigami sample from a 0.002m thin Nylon plastic sheet ($E=9.2$GPa, $\rho=1200$Kg/m$^3$), with cut sizes $l_a=0.04$m, $l_b=0.02$m, $l_c=0.064$m, and cut spacing $w=0.01$m.  Based on the theoretical prediction, this sample provides a Braggs bandgap between 423Hz and 681Hz with a small stretch ($\varphi=33^\circ$, or the kirigami sheet stretched to 183\% of its original length).  Stretching the kirigami sheet to 200\% of its original length ($\varphi=30^\circ$) places the bandgap to between 465 and 773Hz. As the kirigami stretched further to 228\% of its original length ($\varphi=26^\circ$), the bandgap moved to between 550Hz and 838Hz. Such correlation between stretching and bandgap frequencies are confirmed by numerical simulation and experiment. 

While the different test results show good agreement regarding the bandgap frequencies, there are notable differences between the numerical simulation and experiment regarding the magnitude of wave transmissibility within the bandgap.  Such discrepancy comes from damping. The nylon material has significant internal damping, but the numerical simulation assumed zero damping to avoid unnecessary computational complexities.   Regardless, these results firmly validate our assumptions that kirigami cutting and stretching is a simple and effective approach for generating and controlling periodicity in a metamaterial system for wave propagation control. 

\begin{figure}[t]
   \centering
   \includegraphics[]{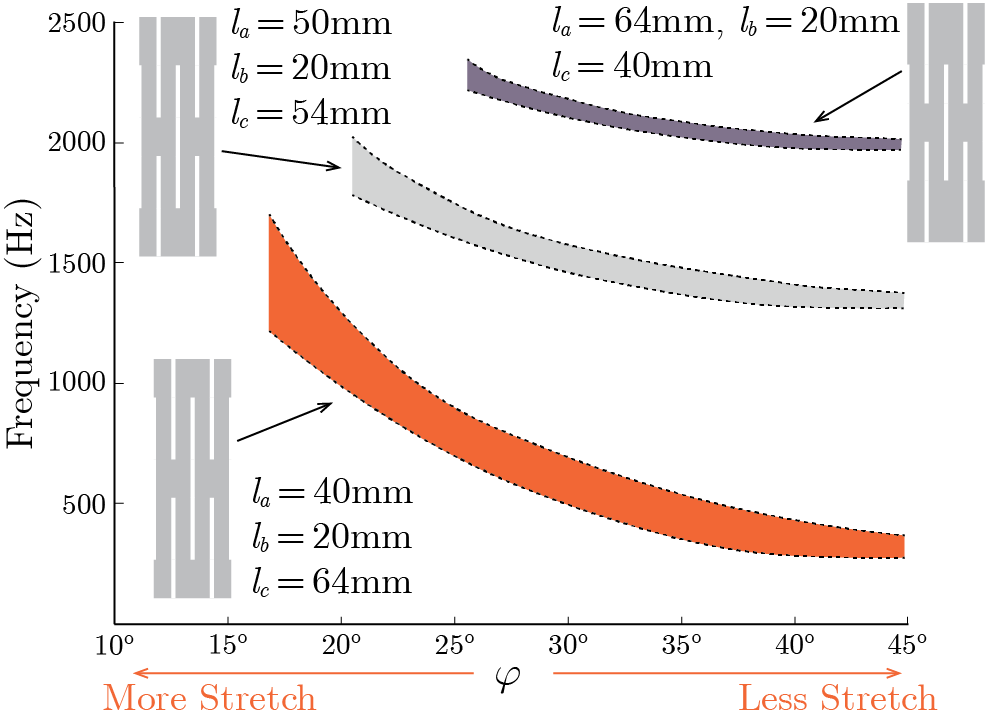}
    \caption{The Braggs bandgap frequencies are directly related to the kirigami cutting pattern design and in-plane stretching. Notice that, the $\varphi$ values correspond to the maximum stretch vary for different cut pattern designs.}
    \label{fig:tuning}
\end{figure}

It is worth pointing out that our analytical calculation and COMSOL simulation use the simplified kirigami geometry with virtual folds and flat facets, but the experiments do not.  That is, the physical kirigami prototype has the curved ligaments. Therefore, the agreement between these results indicates that the kirigami geometry simplification summarized in Figure \ref{fig:geometry} provides an accurate correlation between the Braggs bandgap and kirigami design and stretching.  Therefore, our theory can provide effective design guidelines for kirigami-based metamaterial systems, even with more complicated cut patterns. 

Finally, the simplicity and versatility in kirigami cutting pattern design offer us significant freedom to prescribe and control the wave propagation bandgaps.  We apply our theoretical method to three kirigami sheets and examine their bandgap frequencies at different stretch levels (Figure \ref{fig:tuning}).   These sheets share the same constitutive material, overall size, the cut spacing.  The only difference between them is the cut lengths: which are very simple to change during fabrication.  Yet, the three kirigami sheets exhibit significantly different and tunable bandgap frequencies. 

In conclusion, we use theoretical, numerical, and experimental methods to examine Bragg's elastic wave propagation bandgaps in stretched buckled kirigami sheets.  Our results show that we can prescribe and tune the bandgap frequencies with considerable freedom by simply tailoring the cutting pattern and controlling the in-plane stretch.  Our proposed method for simplifying the kirigami geometry, combined with the plane wave expansion method, provides an accurate and computationally efficient approach for customize-designing the kirigami sheet for various wave and vibration control applications.

\section*{Acknowledgements} \label{sec:acknowledgements}
The authors acknowledge the support from Clemson University (via startup fund and CECAS Dean's Faculty Fellow Award) and National Science Foundation (CMMI-1760943).

\bibliography{Reference.bib}

\appendix*
\section{1. Geometry of Buckled kirigami}
\label{appendix:geometry}
Figure \ref{fig:concept}(a) in the main text explains the parameters defining the parallel and zig-zag distributed cutting pattern.   Here, $l_a$, $l_b$, and $l_c$ define the slit cuts' size; $w$ is the spacing between two cuts along the longitudinal direction.   Once buckled via stretching, the kirigami sheet takes a complicated three-dimensional shape with significant out-of-plane deformation.  As a result, the buckled kirigami structure has a finite thickness, satisfying Euler-Bernoulli's beam geometrical conditions.  In other words, the stretch-buckled kirigami sheet becomes a beam-like structure consisting of a linear periodic array of ``unit cells'' as shown in Figure \ref{fig:concept}(b, c).  

The ligaments inside the unit cells exhibit complex deformations with a non-uniform curvature distribution, making it challenging to calculate the parameters relevant to wave propagation analyses, such as the cross-sectional area and area moment of inertia.  To address this challenge, we introduce ``virtual folds'' at critical locations on the ligaments that exhibit the most concentrated bending deformation and assume the facets in-between these folds are flat surfaces (Surfaces A, B, and C in Figure \ref{fig:geometry} of the main text). The simplified kirigami becomes rigid-foldable, making the kirigami deformation a one degree-of-freedom mechanism. 

\begin{figure}[t]
    \includegraphics[]{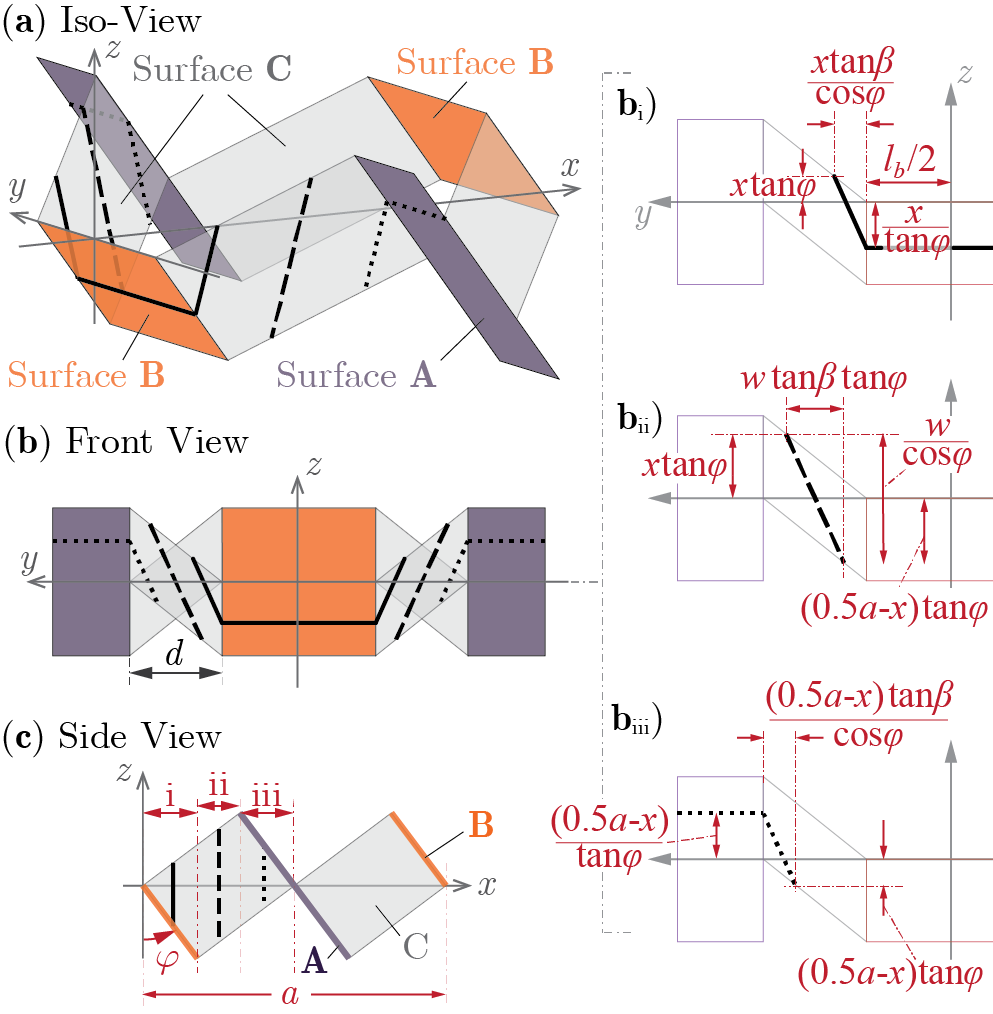}
    \caption{
     The unit cell geometry of a simplified kirigami sheet with a relatively large stretch.  (a-c): The isometric, front, and side view of kirigami unit cell, respectively.  Notice that the independent variable $\varphi$ and the three sections in the half unit cells are highlighted in the side view. (b$_i$-b$_{iii}$): Close-up front view of the cross-section area corresponding to different $x$ values.  Notice that the solid, dashed, dotted black lines are the cross-section of the kirigami unit cell in three different sections.
    }
    \label{fig:geodetail}
\end{figure}

To solve the dynamic equation of motion for wave propagation, we need to calculate the distribution of cross-sectional area $A(x)$ and second moment of inertia $I(x)$ over a unit cell, where $x$ represents the longitudinal direction.   To this end, we choose the dihedral angle ($\varphi$) between the Surface B within the kirigami unit cell and the $y-z$ reference plane as the independent variable (Figure \ref{fig:geometry}(b) and \ref{fig:geodetail}).  When the kirigami sheet is un-deformed (or flat in the $x-y$ reference plane), $\varphi$ takes the maximum value ($\varphi_{max} = \pi/2$); when the buckled kirigami sheet is fully stretched, $\varphi$ take the minimum value:
\begin{equation}
    \varphi_\text{min} = \tan^{-1} \left( \frac{2w}{l_c} \right).
\end{equation}

We also denote the dihedral angle between Surface C defined in Figure \ref{fig:geometry} and the $x-z$ reference place as $\beta$, so that
\begin{equation}
    \beta = \cos^{-1} \left(\frac{2w}{l_c \tan \varphi} \right).
\end{equation}

$\beta = \pi/2$ when the kirigami sheet is un-deformed (flat), and $\beta=0$ when the kirigami sheet is fully stretched. The overall length of the unit cell is
\begin{equation}
    a = \frac{2w}{\sin\varphi}.
\end{equation}

Another important geometric variable is the \emph{projected} length of Surface C along the $y$ axis as illustrated in Figure \ref{fig:geodetail}(b):
\begin{equation}
    d =w\frac{\tan \beta}{\tan \varphi} =\sqrt{\frac{l_c^2}{4}-\frac{w^2}{\tan^2 \varphi}}=\frac{l_c}{2}\sin \beta.
\end{equation}

Here, we assume the kirigami sheet is highly stretched after buckling in that $\phi_{min}<\varphi<\pi/4$ (assuming $l_c>2w$).  We can divide half of the unit cell in this case into three sections (Section i, ii, and iii in Figure \ref{fig:geodetail}(c)) because the cross-sections take distinct shapes in these three sections.

The first Section i corresponds to $0 < x < w\sin \varphi$, and the cross-sectional areas involves surface B and surface C.  The cross section (solid lines in Figure \ref{fig:geodetail}(b$_\text{i}$)) is the summation of these two parts so that
\begin{equation}
    \begin{split}
         A_i(x) & =\frac{t}{\sin \varphi}\left(l_{b}+2L_{i}^C\right) \\ 
                & =\frac{t}{\sin \varphi} \left(l_b + \frac{x}{w \sin \varphi} \sqrt{4w^2+l_c^2 \tan^4 \varphi} \right),
    \end{split}
\end{equation}
where $t$ is the thickness of the kirigami sheet material.  The bending moment of inertia with respect to the $y$-axis also includes two parts:
\begin{equation}
    \begin{split}
    I_i(x) & =I^{B}+l_b\frac{t}{\sin \varphi}\left(\frac{x}{\tan \varphi}\right)^2+2 I_{i}^C +\ldots \\
    & 2\frac{x \tan \beta}{\cos \varphi}\frac{t}{\sin \varphi}\left(\frac{x\tan \varphi - x\cot \varphi}{2}\right)^2,
    \end{split}
\end{equation}
where the first two terms come from the cross-section of Surface B (using parallel axis theorem), and the third and fourth terms come from the cross-section of Surface C. Here, $I^B$ is the bending moment of inertia of the Surface B's cross-section with respect to its own neutral axis in that
\begin{equation}
    I^{B} = \frac{l_b}{12}\left(\frac{t}{\sin \varphi} \right)^3.
\end{equation}

\begin{figure*}[t]
   \centering
    \includegraphics[]{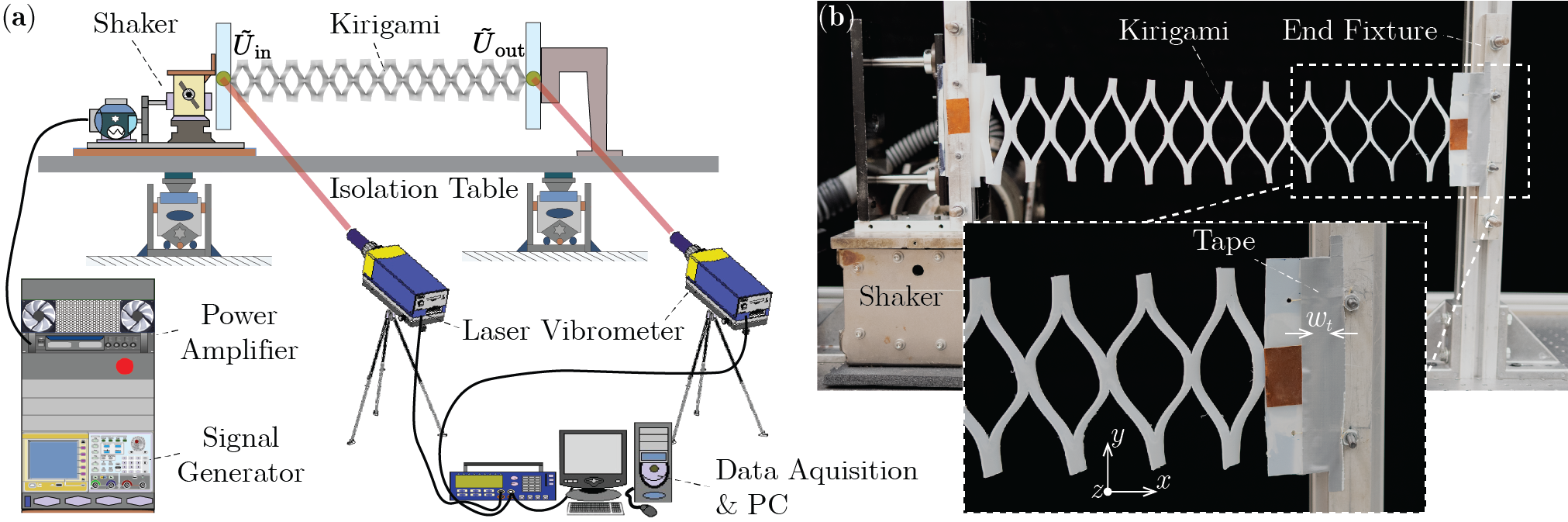}
    \caption{Experimental setup. (a) A schematic diagram of the test apparatus. (b) Nylon-based kirigami prototype on the test stand.}
    \label{fig:ExpSetup}
\end{figure*}

Similarly, $I_i^C$ is Surface C's bending moment of inertia with respect to its own neutral axis in that
\begin{equation}
    I_i^C(x) = \frac{t}{12 \sin \varphi} \frac{x \tan \beta}{\cos \varphi} \left[ \frac{t^2}{\sin^2 \varphi}+ x^2(\tan \varphi+\cot \varphi)^2\right]
\end{equation}

The second Section ii corresponds to $w \sin < x <a/2- w \sin \varphi$, and the overall cross-section area involves Surface C (dashed line in Figure \ref{fig:geodetail}(b$_\text{ii}$)).
\begin{equation}
    A_{ii}(x) = 2 \frac{t}{\sin \varphi} L_{ii}^{C}  =\frac{t}{\sin \varphi} \sqrt{4w^2+l_c^2 \tan^4 \varphi}.
\end{equation}

The bending moment of inertia with respect to the $y$-axis is
\begin{equation}
    I_{ii}(x)  = 2 I_{ii}^{C} + 2 \frac{t}{\sin \varphi}\left[(a/4-x)\tan \varphi \right]^2,
\end{equation}
where
\begin{equation}
    I_{ii}^{C}=\frac{t}{12\sin \varphi} w \tan \varphi \tan \beta \left( \frac{t^2}{\sin^2 \varphi}+\frac{w^2}{\cos^2 \varphi} \right).
\end{equation}

Finally, the third Section iii of the half unit cell corresponds to $\sin \varphi <x< a/2$.  In this section, the cross-section (dotted line in Figure \ref{fig:geodetail}(b$_\text{iii}$)) involve Surface A and C in that
\begin{equation}
    \begin{split}
            A_{iii}(x) & = \frac{t}{\sin \varphi} \left(l_{a}+2 L_{iii}^C \right) \\
                       & =\frac{t}{\sin \varphi}\left(l_a + \frac{a/2-x}{w \sin \varphi} \sqrt{4w^2+l_c^2 \tan^4 \varphi} \right).
    \end{split}
\end{equation}

The corresponding bending moment of inertia with respect to the $y-$axis includes two components in that,  
\begin{equation}
    \begin{split}
    I_{iii}(x) & = I^{A}+l_a\frac{t}{\sin \varphi}\left(\frac{a/2-x}{\tan \varphi}\right)^2 +2 I_{iii}^C + \ldots \\
    & 2\frac{(a/2-x) \tan \beta}{\cos \varphi}\frac{t}{\sin \varphi}\left[\frac{(a/2-x)(\tan \varphi - \cot \varphi)}{2}\right]^2,
    \end{split}
\end{equation}
where, 
\begin{equation}
\begin{split}
     I_{iii}^C(x) = & \frac{t}{12 \sin \varphi} \frac{(a/2-x) \tan \beta}{\cos \varphi} \ldots \\
                    & \left[ \frac{t^2}{\sin^2 \varphi}+ (a/2-x)^2(\tan \varphi+\cot \varphi)^2\right],
\end{split}
\end{equation}
and $I^A$ is the bending moment of inertia of the cross-section area in Surface A with respect to its own neutral axis so that
\begin{equation}
    I^{A} = \frac{l_a}{12}\left(\frac{t}{\sin \varphi} \right)^3.
\end{equation}

\section{2. Simulation and Experiment Methods}

We use the numerical analysis and experiment results to validate the predicted Bragg's bandgap from our theory.  Regarding numerical analysis, we first create a CAD model of stretch-buckled kirigami sheet in SolidWorks --- based on the simplified geometry shown in Figure \ref{fig:geometry} --- and export it to COMSOL software.   The constitutive material properties of the kirigami sheet, including elastic modulus and mass density, are set consistent with the Nylon-based experiment sample (except for damping, as we discuss in the main text).   Then, we apply fine mesh to guarantee the convergence of simulation results and analyze the kirigami's frequency response.   That is, we use two reference lines parallel to the $y-$direction at both ends of the kirigami sheet to measure the vibration entering and exiting structure at specific ranges of frequencies.  The transmissibility of the kirigami sheet is
\begin{equation}\label{ff}
    TR = 20 \log \frac{\tilde{U}_\text{out}}{\tilde{U}_\text{in}},
\end{equation}
where $\tilde{U}_\text{in}$ is the maximum displacement of entering elastic wave, and $\tilde{U}_\text{out}$ is the maximum measured displacement of transmitted vibration.

Figure \ref{fig:ExpSetup} shows the experiment setup. We attach one end of the Nylon-based kirigami sheet to a shaker (Labworks DB-140 with Pa-141 amplifier), which provides out-of-plane ($z-$direction) excitation.   The other end of the kirigami sheet is attached to a rigid end fixture via thin tape.  In this way, the kirigami is ``free'' to vibrate in the $z-$direction but constrained in the in-plane $y-$direction, preventing the occurrence of twisting motion.  According to the excitation amplitude, we carefully choose the tape width $w_t$ (or the distance between the kirigami sheet's free end and rigid fixture) to mimic the free boundary condition the best.   The tape needs to be wide at low excitation frequency due to a large vibration amplitude but narrow at high frequency with a smaller amplitude.   We found that $\tan^{-1}(w_t/\tilde{U}_\text{out})<6^{^\circ}$ is a good guideline for tape width selection.  To mitigate external sources of noises like the air conditioner airflow, especially at the higher frequencies, we place the test apparatus on the isolation table (Newport RS-2000) and cover it with acrylic sheets.  

To measure the wave transmissibility through the stretch-buckled kirigami sheet, we use a signal generator (Tektronix AFG3022c) to generate harmonic excitations with sweeping frequencies from 0 to 1000Hz and use two laser vibrometers (Polytec OFC-5000) to measure the input and transmitted wave (Figure \ref{fig:ExpSetup}).   Fourier transformation with some smoothing (to reduce the electronic noises) generates the results shown in Figure \ref{fig:result}(c).

\end{document}